\documentclass[aps,amsmath,amssymbd,dvips,preprint,nofootbib,showkeys,showpacs,preprintnumbers]{revtex4}

\usepackage{graphicx}
\usepackage{color}

\def \be  {\begin{equation}}
\def \ee  {\end{equation}}
\def \ee  {\end{equation}}
\def \bea {\begin{eqnarray}}
\def \eea {\end{eqnarray}}

\begin{document}

\preprint{ECTP-2015-08}
\preprint{WLCAPP-2015-08}
\vspace*{3mm}

\title{QCD thermodynamics and magnetization in nonzero magnetic field}

\author{Abdel Nasser Tawfik\footnote{The authors declare that there is no conflict of interest regarding the publication of this manuscript.}}
\email{a.tawfik@eng.mti.edu.eg}
\affiliation{Egyptian Center for Theoretical Physics (ECTP), Modern University for Technology and Information (MTI), 11571 Cairo, Egypt}
\affiliation{World Laboratory for Cosmology And Particle Physics (WLCAPP), Cairo, Egypt}

\author{Asmaa G. Shalaby}
\affiliation{Physics Department, Benha University, 13815 Benha, Egypt}
\affiliation{World Laboratory for Cosmology And Particle Physics (WLCAPP), Cairo, Egypt}

\author{Abdel Magied Diab}
\affiliation{Egyptian Center for Theoretical Physics (ECTP), Modern University for Technology and Information (MTI), 11571 Cairo, Egypt}
\affiliation{World Laboratory for Cosmology And Particle Physics (WLCAPP), Cairo, Egypt}

\author{Nada Ezzelarab}
\affiliation{Egyptian Center for Theoretical Physics (ECTP), Modern University for Technology and Information (MTI), 11571 Cairo, Egypt}
\affiliation{World Laboratory for Cosmology And Particle Physics (WLCAPP), Cairo, Egypt}

\begin{abstract} 
In nonzero magnetic field, the magnetic properties and  thermodynamics of the quantum-chromodynamic (QCD) matter is studied in the hadron resonance gas and the Polyakov linear-sigma models and compared with recent lattice calculations. Both models are fairly suited to describe the degrees of freedom in the hadronic phase. The partonic ones are only accessible by the second model. It is found that the QCD matter has paramagnetic properties, which monotonically depend on the temperature and are not affected by the hadron-quark phase-transition. Furthermore, raising the magnetic field strength increases the thermodynamic quantities, especially in the hadronic phase but reduces the critical temperature, i.e. inverse magnetic catalysis. 
\end{abstract}
\date{\today}


\maketitle
\tableofcontents
\makeatletter
\let\toc@pre\relax
\let\toc@post\relax
\makeatother 

\section{Introduction}
\label{sec:intr}

In peripheral heavy-ion collisions, a strong and very localized magnetic field is likely generated. The opposite relativistic motion of the spectator's positive charges and the imbalance in the momentum of the participants together contribute to generating such short-lived and huge magnetic field ($B\sim 10^{19}$ Gauss), which apparently should come up with significant effects on the quantum chromodynamic (QCD) matter. These effects  can be coupled to experimental observables, for instance, in the STAR experiment at the relativistic heavy-ion collider (RHIC), $|q| B \sim m_{\pi}^2$ \cite{Skokov:2009, Bzdak:2012}, and in ALICE experiment at the large hadron collider (LHC), $|q| B \sim 10-15 m_{\pi}^2$ \cite{Skokov:2009,Bzdak:2012}, where $|q|$ is the net electric charge, and $m_{\pi}$ is the pion mass \cite{Jens11}.  Only in heavy-ion collisions do the self-generating magnetic fields play an essential role. The early Universe, and magnetors (special types of neutron stars), for instance, should extremely be affected by such fields, which are conjectured to influence even the acceleration of various cosmic rays and the creation of stars  \cite{Lazarian09}. They can mediate important processes affecting the dynamics, the distribution, and even the composition of the galactic plasmas, for instance \cite{Lazarian09}.

The influences of magnetic fields on the hadronic matter and on the phase-space structure of quark-gluon plasma (QGP) are included in various models, such as hadron resonance gas (HRG) \cite{HRG1,HRG2}, and estimated in  lattice QCD simulations \cite{QCD:2012,lattice:2013b, lattice:2013,QCD:2013c,QCD:2013d,lattice:2014}. The Polyakon Nambu-Jona-Lasinio (PNJL) and NJL models  are examples on QCD-like models in which such magnetic effects were estimated \cite{Klevansky:1992, NJLsu2, Menezes:2009, Fukushima:2010l}. Coupling Polyakov loops to the linear sigma model (PLSM) introduces color charge interactions to the pure gauge field. PLSM reveals interesting features about the response of QCD matter to finite magnetic field \cite{TN:magnet, TN:magnet2}. 
Recently, electric conductivity in thermal medium  and the phase structure of the strongly interacting matter in presence of magnetic field have been reported \cite{review1,review2,review3,Shovkovy:2013}.

In the present work, we plan to utilize the HRG and PLSM approaches in finite magnetic field  in order to study the QCD equation-of-state (EoS). Furthermore, it intends to estimate different thermodynamic observables including pressure, entropy, energy densities and magnetization by using the modified energy-momentum dispersion relations which arise from finite magnetic field. Also, we verify that the thermal QCD medium is paramagnetic, especially at the critical temperature. Our calculations are confronted to recent lattice QCD simulations \cite{lattice:2014}. A quite good agreement is observed.

The present paper is organized as follows. We summarize the fundamentals of Landau quantizations in section \ref{sec:landau}. The HRG model in nonzero magnetic field is discussed in section \ref{Sec:EoSHRG}. PLSM is briefly introduced in section \ref{Sec:PLSM}. The obtained  results are confronted to recent lattice QCD  in section \ref{Sec:Res}. The conclusions are outlined in section \ref{Sec:con}.

\section{Formalism}

\subsection{General Remarks on Landau Quantization} 
\label{sec:landau}

In nonzero magnetic field, the eigen-energy can be given as,
\begin{equation}
E = \left(n+\frac{1}{2}\hbar\, \omega_{H}\right)+\frac{p_{z}^{2}}{2\, m}-\mu \frac{s_{z}}{s} \, H, \label{Landau}
\end{equation}
where the angular frequency $\omega_{H}=|e| H/m c$ and discrete values with integer quantum number $n=0,1,2,\cdots$ called Landau levels are assigned to the first term \cite{book2}. This describes the motion in a plane perpendicular to the magnetic field, which is directed towards $z$-axis. Accordingly, the dispersion relation should be modified \cite{Fraga:2008, book2} 
\begin{equation} 
E_{\ell s_{z}}= \left[p_{z}^{2}+m^{2}+2 |q| (n-S_{z}+1/2) B \right]^{1/2}, \label{disp1}
\end{equation}
where the $S_z$ is the component of the spin in $z$-direction, $n$ being the index to label the Landau levels and $|q|>0$ is the electric charge of $i$-th hadron. In PLSM, the dispersion relation also becomes modified 
\begin{eqnarray}
E_{B, f} (B) =\left[p_{z}^{2}+m_f^{2}+|q_f|(2n+1-\sigma) B\right]^{1/2}, \label{eqLL}
\end{eqnarray}
where the quantization number ($n$) known as the Landau quantum number, $\sigma$ is related to the spin quantum number, $\sigma=\pm S_z/2$ and the $q_f$ ($m_f$) being quark electric charge (mass). The quark masses are directly coupled to the $\sigma$-fields through Yukawa coupling $g$ as  
\bea
m_l &=& g\, \frac{\sigma_l}{2}, \\ 
m_s &=& g\, \frac{\sigma_s}{\sqrt{2}},  \label{qmassSigma}
\eea
where the subscript $l$ refers to degenerate light up and down quarks.

It is worthwhile to notice that $2n+1-\sigma$ can be replaced by sum over the Landau Levels $0\, \leq \nu \, \leq \nu_{max_f}$. The lower value refers to the Lowest Landau Level (LLL), while the higher one stands for the Maximum Landau Level (MLL), $\nu_{max}$, which contributes to the maximum quantization number ($\nu_{max_{f}} \rightarrow \infty$). For sake for completeness, we mention that $2-\delta_{0 \nu}$ represents degenerate Landau Levels.

\subsection{Hadron resonance gas model in nonzero magnetic field}
\label{Sec:EoSHRG}
 
Treating hadron resonances as a collision-free gas \cite{Karsch:2003vd,Karsch:2003zq,Redlich:2004gp,Tawfik:2004sw,Taw3} allows the estimation of thermodynamic partition function. For a recent review, the readers can consult Ref. \cite{Tawfik:2014eba}. Even for an interacting system, such ideal gas, which is composed of hadron resonances with masses $\le 2~$GeV~\cite{Tawfik:2004sw,Vunog}, where the inclusion of heavy resonances effectively introduces interactions and correlations to the system, can be described by grand canonical partition function that gives a quite satisfactory description for the particle production in heavy-ion collisions and the lattice QCD thermodynamics \cite{Karsch:2003vd,Karsch:2003zq,Redlich:2004gp,Taw3b,Taw3c}
\bea
\ln Z(T,V,\mu) &=& \sum_i\pm \frac{V d_i}{2\pi^2} \int_0^{\infty} p^2 \ln\left[1\pm \, \exp[(\mu_i -E_i)/T]\right]\, d p, \label{eq:PFq1}
\eea
where $\pm$ stands for fermions and bosons, respectively, and $V$ is the volume and $d_{i}$ is the degeneracy for $i$-th hadron.  As discussed earlier, the dispersion relation ($E_{i}$) for charged hadrons in nonzero magnetic field is a subject of modification, Eq. (\ref{disp1}), while for neutral hadrons, $E_{i}= \sqrt{p^{2}+m_{i}^{2}}$. 

In finite magnetic filed, the phase space integral in Eq. (\ref{eq:PFq1}) is to be expressed as a one-dimensional integral (dimension reduction). The partition function for $i$th particle in presence of magnetic field can be written as,
\bea
\ln Z(T,V,\mu) &=& \pm \frac{V d_i}{2\pi^2}|q_{i}| e B_z \sum_n \sum_{S_z} \int_{0} ^{\infty}  \ln\left[1 \pm \exp\left(\frac{\mu_i -E_{i}}{T}\right)\right]  d p_z. \label{PartiHRGeB}
\eea
where $p_z$ is the component of the particle momentum along the direction of the magnetic field. The spin in $z$-direction ($s_{z}$) is running as $-S_z, -S_z+1,\cdots+S_z$, where $S_z$ is the resonance spin in $z$-direction. For $i$-th particle, the pressure, energy density and entropy, respectively, read
\bea
P_{i} &=& \pm \frac{d_{i} T}{2\, \pi^{2}} |q_{i}| eB \sum_{\kappa} \sum_{s_{z}} \int _{0}^{\infty}  \ln\left[1 \pm \exp\left(\frac{\mu_i -E_{i}}{T}\right)\right]\, d p_{z}, \label{Eq:presChT}  \\
\varepsilon_{i} &=& \pm \frac{d_{i} T}{2\pi^{2}} |q_{i}| eB \sum_{\kappa} \sum_{s_{z}} \int _{0}^{\infty}  \frac{E_{i}}{ \exp[(E_{i}-\mu_i)/T]\pm 1}\, d p_{z}, \label{eq:s_Ez} \\
s_i &=& \pm \frac{d_i}{2 \pi^2} |q_i| eB \sum_{\kappa} \sum_{s_z} \int_{0}^{\infty}  \left\{
 \ln\left[1 \pm \exp\left(\frac{\mu_i-E_i}{T}\right)\right] \pm \frac{(E_i-\mu_i)/T}{\exp[(E_i-\mu_i)/T] \pm 1} \right\}\, d p_{z}.
\label{eq:s_Bz}
\eea
The magnetization, Eq. (\ref{eq:Mm}), is a vector field indicating  the creation of magnetic dipole moments resulting from the response of the material to nonzero magnetic field.
 
The dependence of hadron masses on nonzero magnetic field still questionable \cite{Taya:2014}. In LHC, the magnetic field of noncentral heavy-ion collisions can reach up to $10-15\, m^{2}_{\pi}$. This value is almost identical to $\Lambda_{QCD}$ \cite{Skokov:2009,Luschevskaya:2014}. At high momentum transfer, i.e. asymptotic freedom, the QCD strength becomes very small at short distances, and the leading order running strong coupling  reads  \cite{Igor:2013} 
\begin{equation}
\frac{1}{\alpha_{s}} \simeq \beta_{0} \ln \left[ \frac{|eB|}{\Lambda_{QCD}^{2}}\right],
\end{equation}
where $\beta_0 = (11N_c - 2N_f)/12\pi$ and the QCD phase transition might take place at $\Lambda_{QCD} \sim 0.2~$GeV or $B \sim 0.2~$GeV$^{2}$. In this respect, the magnetic field can be categorized into
\begin{itemize}
\item $ |eB| \gg \Lambda_{QCD}^{2}$, strong and
\item $ |eB| \ll \Lambda_{QCD}^{2}$, weak magnetic field.
\end{itemize}

On the other hand, in a finite magnetic field, the lattice QCD calculations refer to mass hierarchy \cite{Luschevskaya:2014}. It has been found that, $m_{\rho^0} \sim m_{\pi^+} > m_{ \rho^+} \sim m_{\pi^0}$ at $e B \gg \Lambda_{\rm QCD}^2$  \cite{Taya:2014}. 
Thus, the strong magnetic field is the one at which $\Lambda_{QCD}^{2}\ll |eB| \lesssim (10~$TeV$)^{2} $. This means that the value of $\Lambda_{QCD}^{2}$ is small to contribute to the dynamical quark masses or the hadron constituents, i.e. QCD represents an  intermediate regime \cite{Luschevskaya:2014}.

\subsection{Polyakov linear-sigma model in nonzero magnetic field}
\label{Sec:PLSM}

In SU(3)$_L\times$ SU(3)$_R$ symmetries,  the Lagrangian of LSM for $N_f=3$ lavors ($u$,  $d$ and $s$-quarks) with $N_c=3$ color degrees of freedom is given as 
\bea 
\mathcal{L}= \mathcal{L}_{\mathrm{chiral}} -\mathcal{U}(\phi,\, \phi^*,\,T), \label{plsm}
\eea
where the chiral Lagrangian consists of two parts; fermionic  and  mesonic. Both couple to each other with a flavor-blind Yukawa coupling constant $g$ \cite{blind}. 
\bea
\mathcal{L}_q &=&\sum_{f} \,\bar{\psi_f} \left[ i \gamma^\mu D_{\mu} - g T_a (\sigma_a \,+\, i \gamma_5 \pi_a)\,\right] \psi_f ,  \label{lfermion} \\
\mathcal{L}_m &=&
\mathrm{Tr}(\partial_{\mu}\Phi^{\dag}\partial^{\mu}\Phi-m^2
\Phi^{\dag} \Phi)-\lambda_1 [\mathrm{Tr}(\Phi^{\dag} \Phi)]^2 
-\lambda_2 \mathrm{Tr}(\Phi^{\dag}
\Phi)^2+c[\mathrm{Det}(\Phi)+\mathrm{Det}(\Phi^{\dag})]
+\mathrm{Tr}[H(\Phi+\Phi^{\dag})], \hspace*{7mm} \label{lmeson} 
\eea
where $T_a=\lambda_a/2$ with $a=0,1, \cdots, 8$ are the nine generators of the $U(3)$ symmetry group and $\lambda_a$ are the eight Gell-Mann matrices \cite{Gell Mann:1960}. $\sigma_a$ are the scalar and $\pi_a$ are the pseudoscalar mesons,   
\begin{eqnarray}
\Phi= T_a \phi _{a} =T_a(\sigma_a+i\pi_a).\label{Phi}
\end{eqnarray}
The spontaneous and explicit chiral symmetry breaking is given in more details in Ref. \cite{Gasiorowicz:1969,TawfikPLSM1,LSMMasses:2014}. 

The second term in Eq. (\ref{plsm}), $\mathbf{\mathcal{U}}(\phi, \phi^*, T)$, represents the Polyakov-loop effective potential \cite{Polyakov:1978vu}, where the Polyakov loop fields, $\phi$ and $\phi^{*}$ are the order parameters for deconfinement phase-transition  \cite{Ratti:2005jh,Schaefer:2007d}. In the present work, we implement a polynomial expansion in $\phi$ and $\phi^{*}$ \cite{Ratti:2005jh,Roessner:2007,Schaefer:2007d,Fukushima:2008wg}
\bea
\frac{{\bf \mathcal{U}}(\phi,\, \phi^*, T)}{T^4}=-\frac{b_2(T)}{2}|\phi|^2-\frac{b_3}{6}(\phi^3+\phi^{*3})+\frac{b_4}{4}(|\phi^2|)^2, \label{Uloop}
\eea
where $b_2(T)=a_0+a_1\left(T_0/T\right)+a_2\left(T_0/T\right)^2+a_3\left(T_0/T\right)^3$ with $a_0=6. 75$, $a_1=-1. 95$, $a_2=2. 625$, $a_3=-7. 44 $, $b_3 = 0.75$ and $b_4=7.5$ \cite{Ratti:2005jh}. The deconfinement temperature $T_0$ is fixed at $270~$MeV.  

In the mean field approximation, the thermodynamic potential estimates the energy exchange between quarks and antiquarks at temperature ($T$) and baryon chemical potential ($\mu_f$), where $f$ runs over the quark flavors. The thermodynamic potential, $\Omega=-T \, \ln\mathcal{Z}/V$, reads
\begin{equation}
\Omega =  U(\sigma_l, \sigma_s)+\mathbf{\mathcal{U}}(\phi, \phi^*, T)+\Omega_{\bar{q}q}, \label{potential}
\end{equation}
where the first term represents the pure mesonic part
 \bea
U(\sigma_l, \sigma_s) &=& - h_l \sigma_l - h_s \sigma_s + \frac{m^2}{2}\, (\sigma^2_l+\sigma^2_s) - \frac{c}{2\sqrt{2}} \sigma^2_l \sigma_s   
+ \frac{\lambda_1}{2} \, \sigma^2_l \sigma^2_s +\frac{(2 \lambda_1 +\lambda_2)}{8} \sigma^4_l + \frac{(\lambda_1+\lambda_2)}{4}\sigma^4_s. \hspace*{8mm} \label{Upotio}
\label{pure:meson}
\eea
This potential can be constructed from $\sigma_x$ for light flavors ($u$- and $d$-quark) and $\sigma_y$ for $s$-quark.  

In finite  magnetic field  (and finite $T$ and $\mu_f$), and by means of Landau quantization and magnetic catalysis, the quark-antiquark potential is given as
\bea 
\Omega_{\bar{q}q}(T, \mu _f, B) &=& - 2 \sum_{f=l,s} \frac{|q_f| B \, T}{(2 \pi)^2} \,  \sum_{\nu = 0}^{\infty}  (2-\delta _{0 \nu })  \int_0^{\infty} d p_z \nonumber \\ && \hspace*{5mm} 
\left\{ \ln \left[ 1+3\left(\phi+\phi^* e^{-\frac{E_{B, f} -\mu _f}{T}}\right)\; e^{-\frac{E_{B, f} -\mu _f}{T}} +e^{-3 \frac{E_{B, f} -\mu _f}{T}}\right] \right. \nonumber \\ 
&& \hspace*{3.5mm} \left.+\ln \left[ 1+3\left(\phi^*+\phi e^{-\frac{E_{B, f} +\mu _f}{T}}\right)\; e^{-\frac{E_{B, f} +\mu _f}{T}}+e^{-3 \frac{E_{B, f} +\mu _f}{T}}\right] \right\}. \label{PloykovPLSM}
\eea
Finally, we should assume global minimization of the thermodynamic potential, 
\begin{eqnarray}
\left.\frac{\partial \Omega}{\partial \sigma_l} = \frac{\partial
\Omega}{\partial \sigma_s}= \frac{\partial \Omega}{\partial
\phi}= \frac{\partial \Omega}{\partial \phi^*}\right|_{min} &=& 0, \label{cond1}
\end{eqnarray}
in order to fix the remaining parameters $\sigma_l=\bar{\sigma_l}$, $\sigma_s=\bar{\sigma_s}$, $\phi=\bar{\phi}$ and $\phi^*=\bar{\phi^*}$ and their dependencies on  $T$, $\mu_f$ and $e B$ \cite{THD:magnetic,TN:magnet,TN:magnet2}.

\section{Results and Discussion}
\label{Sec:Res}

Based on the remarkable success of the HRG model in reproducing lattice QCD thermodynamics, for instance, it is straightforward to perform similar analysis in nonzero magnetic field \cite{QCD:2013c,HRG2}. In peripheral heavy-ion collisions, the magnetic field can be as much as $eB=3.25~$GeV$^{2}$ \cite{lattice2}. We also utilize PLSM in studying various QCD properties and phenomena \cite{lattice:2014}. The magnetization and different thermodynamic quantities are calculated from both models at $eB=0.0-0.4~$ GeV$^{2}$ and compared with recent lattice QCD calculations \cite{lattice:2014}.

\begin{figure}[htb]
\centering{
\includegraphics[width=5.cm,angle=-90]{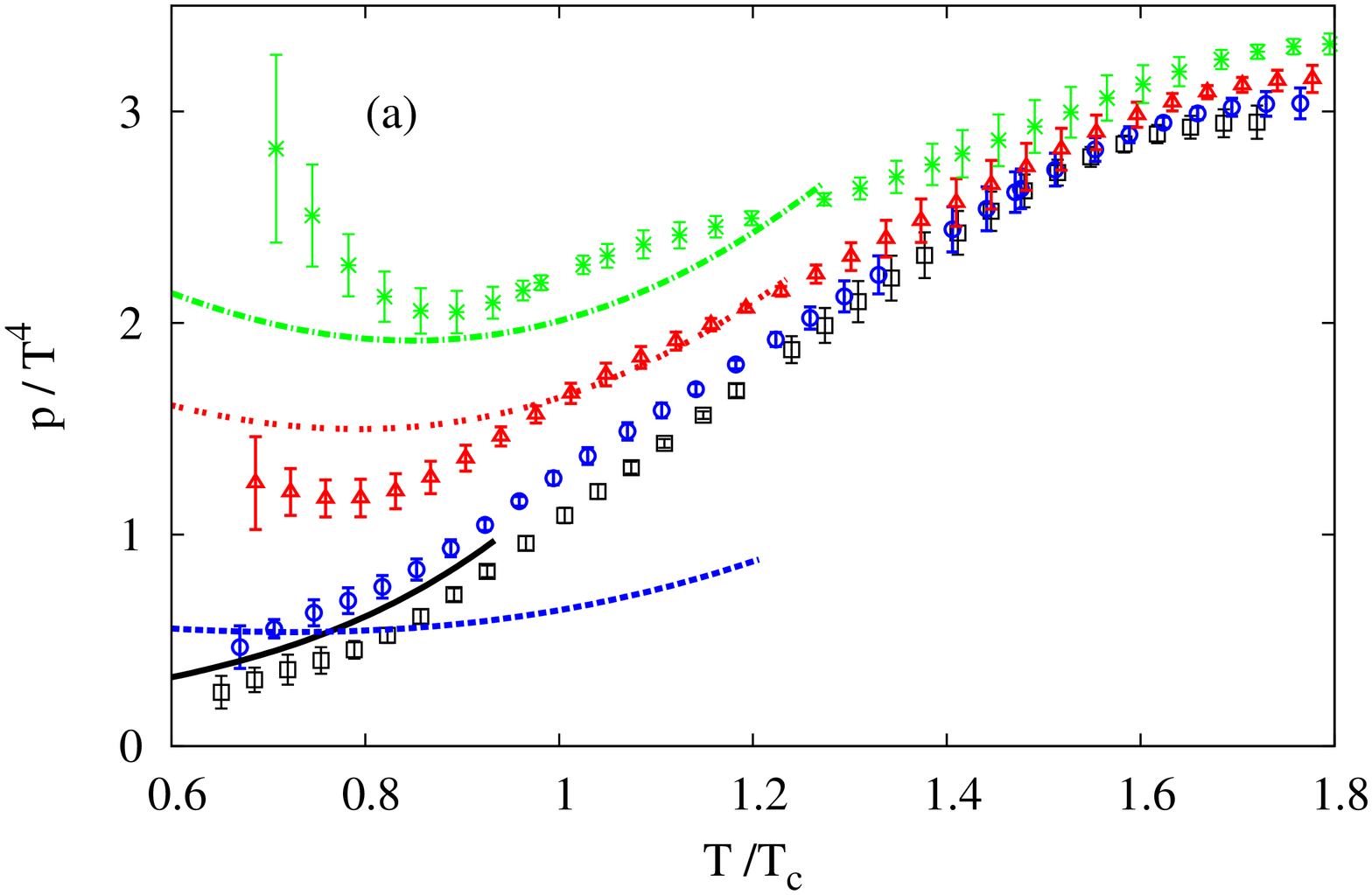}
\includegraphics[width=5.cm,angle=-90]{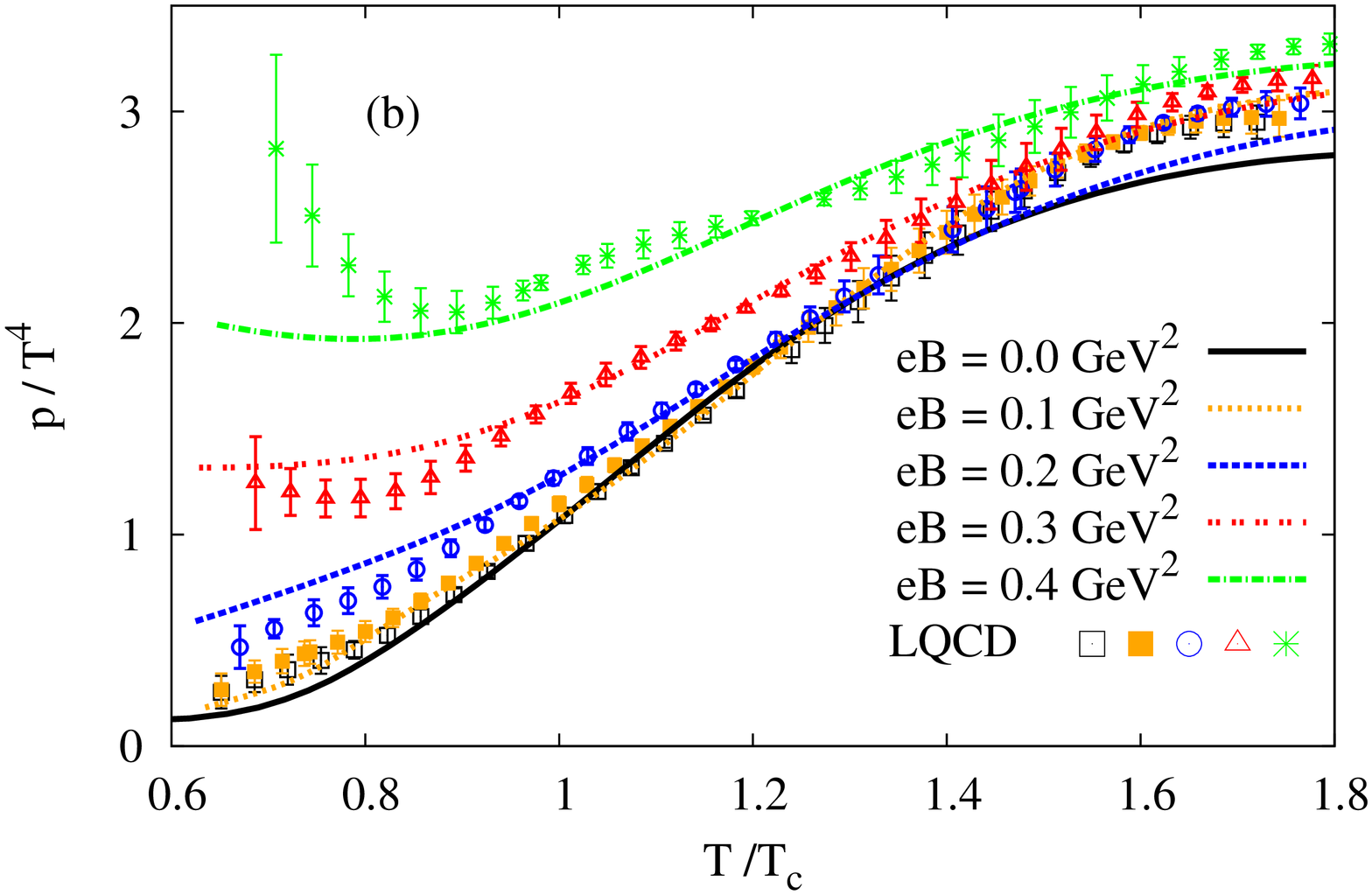}
\caption{Left-hand panel (a): The dependence of normalized pressure on temperature calculated from HRG model, Eq. (\ref{Eq:presChT}) at different values of magnetic fields,  $eB=0.0$, $0.1$, $0.2$, $0.3$ and $0.4$ GeV$^{2}$ compared with recent lattice calculations \cite{lattice:2014} (symbols). Right-hand panel (b): the same as in left-hand panel but the calculations from PLSM, Eq. (\ref{PloykovPLSM}). \label{fig:pressure}}}
\end{figure}

Fig. \ref{fig:pressure} presents the normalized pressure $p/T^{4}$ as function of temperature at $eB=0.0$ (solid), $0.2$ (dashed), $0.3$ (double-dot-dhashed), and $0.4$ GeV$^{2}$ (dot-dashed curve). The results are compared with recent lattice calculations \cite{lattice:2014} at  $eB=0.0$ ( open square), $eB=0.1$ (closed square), $0.2$ (circle), $0.3$ (triangle) and $0.4$ GeV$^{2}$ (asterisk). (a) shows the calculations from the HRG model (curves), while the PLSM calculations are depicted in the right-hand panel. It is apparent that the pressure increases with increasing magnetic field, especially at low temperatures. At high temperatures, $p/T^4$ is limited to the Stefan-Boltzmann limits, which apparently is not affected by the magnetic field.
 
In left-hand panel, we notice that the pressure calculated from HRG slightly deviates from the lattice calculations, although the temperature dependence is preserved. For HRG, $p(T, eB)$ is calculated from Eq. (\ref{Eq:presChT}) without corrections due to lattice $\phi$ and $B$ schemes \cite{lattice:2014}. The right panel of Fig. \ref{fig:pressure} shows the same as in the left-hand panel but for calculations from PLSM. Our results agree well with the lattice simulations \cite{lattice:2014} at a wide range of temperatures. The pressure is directly deduced from PLSM potential, Eq. (\ref{PloykovPLSM}), in mean field approximation, i.e. no corrections with respect to the magnetic flux ($\phi=e B \cdot L_x L_y$) and/or the magnetic field ($B$) have been done in lattice simulations, respectively.

It is obvious that the HRG calculations at low magnetic fields ($\lesssim 0.2~$GeV$^2$) disagree with the corresponding lattice simulations. There are almost temperature independent, which contradicts even the same calculations in vanishing magnetic field and becomes radical at large temperatures (near the critical temperature). It worthwhile to recall that in vanishing magnetic field, the partition function is given by integral over six-dimensional phase-space and the dispersion relations follow Lorentz invariance principle. But, in finite magnetic field, the integral dimensionality is drastically reduced and simultaneously accompanied by a considerable modification in the dispersion relation.  

In finite magnetic field $B$, the velocity of a test particle with momentum $P$ can be deduced from the dispersion relations, Eq. (\ref{disp1}) and Eq. (\ref{eqLL}), respectively, ($v=\partial \epsilon^{total}/\partial P$)
\bea
v_p &=&  c \left[\frac{c\, p}{c\, p + 2 |q| (\kappa+\frac{1}{2}-s_{z}) B}\right], \quad \textrm{for HRG}, \\
v_p &=&  c \left[\frac{c\, p}{c\, p + 2 |q_f|(\kappa+\frac{1}{2}-\frac{\sigma}{2}) B}\right], \quad \textrm{for PLSM},
\eea
where $\sigma=\pm S/2$. 
Then the causality is guaranteed for $v_p$ not exceeding the speed of light $c$, i.e. as long as as the $B$-term is finite positive, which should be estimated, quantitatively, as function of temperature and magnetic field strength. This might give an explanation why HRG fails to reproduce lattice pressure at low magnetic field, while PLSM does not.

For a reliable comparison with the lattice calculations \cite{lattice:2014}, an additional constant magnetic term should be added to the free energy, so that \cite{HRG1}
\begin{eqnarray}
{\cal M} = -\frac{1}{V} T \frac{\partial \ln \mathcal{Z}}{\partial (e B)}     \label{eq:Mm}
\end{eqnarray}
Accordingly, one should distinguish between two setups where either magnetic field $B$ ($B$-scheme) or magnetic flux $\Phi$ is kept constant during the expansion \cite{lattice:2014}, where $L_x$ and $L_y$ are system extensions in $x$- and $y$-direction, respectively. The response of QCD matter (hadrons and partons) to nonzero magnetic field can be estimated from the magnetization. 

In $B$-scheme, the pressure $p$ becomes isotropic but anisotropic in $\phi$-scheme. The compressing force in $\phi$-scheme is directed oppositely to the magnetic field. The longitudinal pressure $p_z$ does not depend on the scheme. The good agreement with lattice calculations, in Fig. \ref{fig:pressure}, especially at large magnetic fields, is obtained where the calculations are simply {uncorrected} or scheme-independent free-energies are assumed.

\begin{figure}[htb]
\centering{
\includegraphics[width=5.cm,angle=-90]{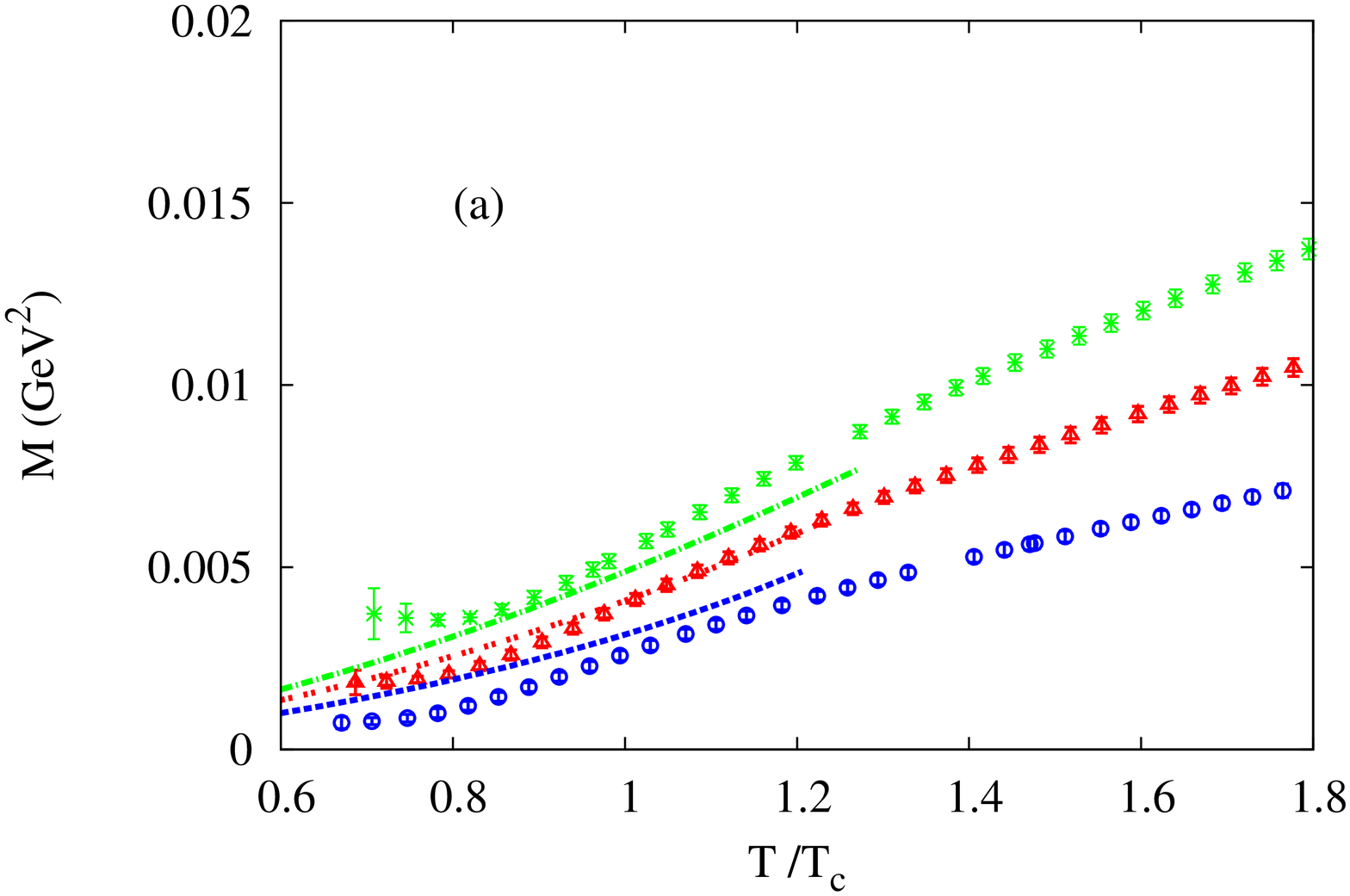}
\includegraphics[width=5.cm,angle=-90]{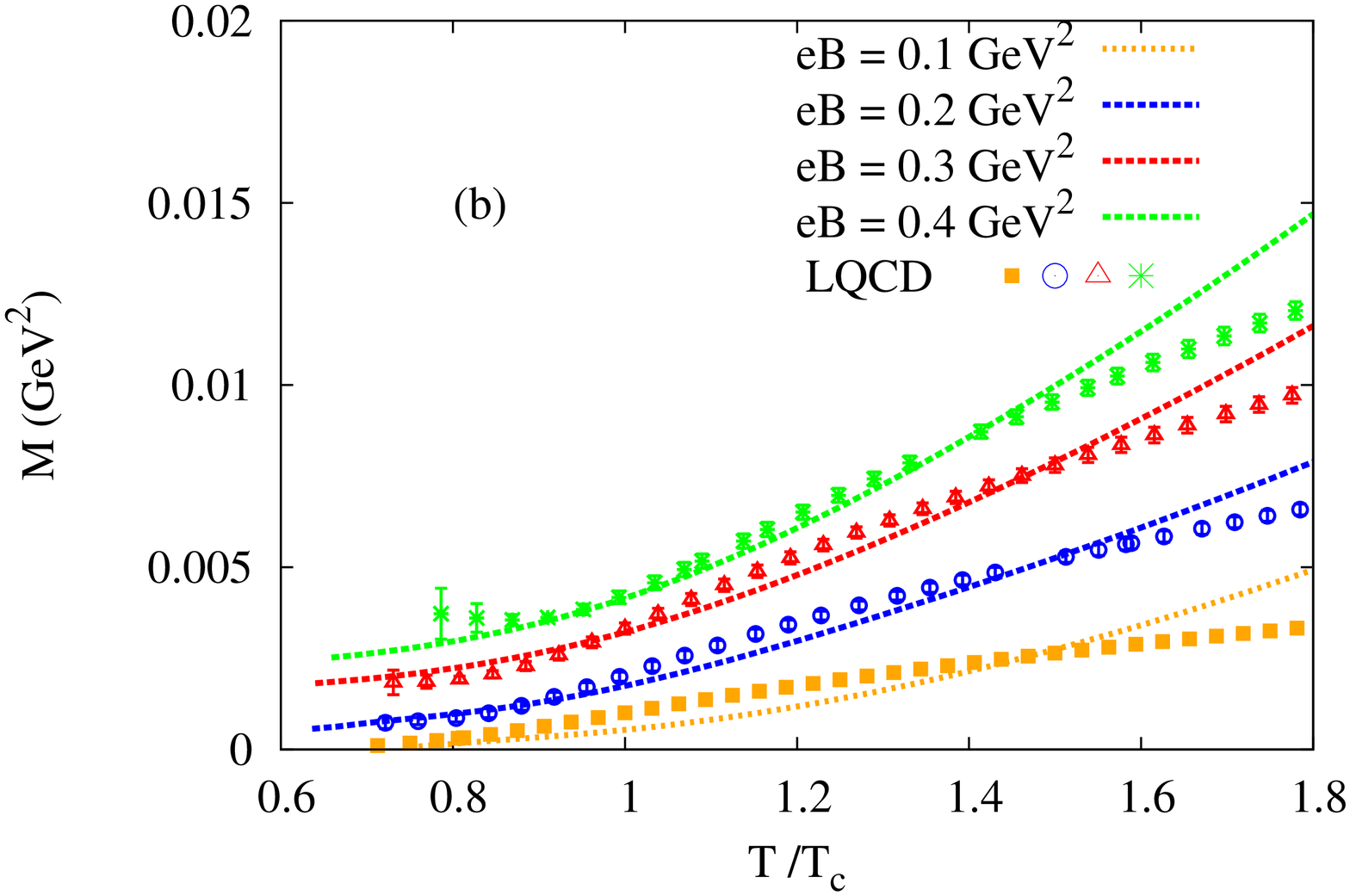}
\caption{(Online color) The magnetization ${\cal M}$ is given as function of temperature at $eB=0.0-0.4~$GeV$^{2}$ and compared with lattice results (symbols) \cite{lattice:2014}. The HRG results are depicted in the left-hand panel, while that from PLSM are in the right-hand panel} \label{fig:Magnetization}
}
\end{figure}

\begin{figure}[htb]
\centering{
\includegraphics[width=5.cm,angle=-90]{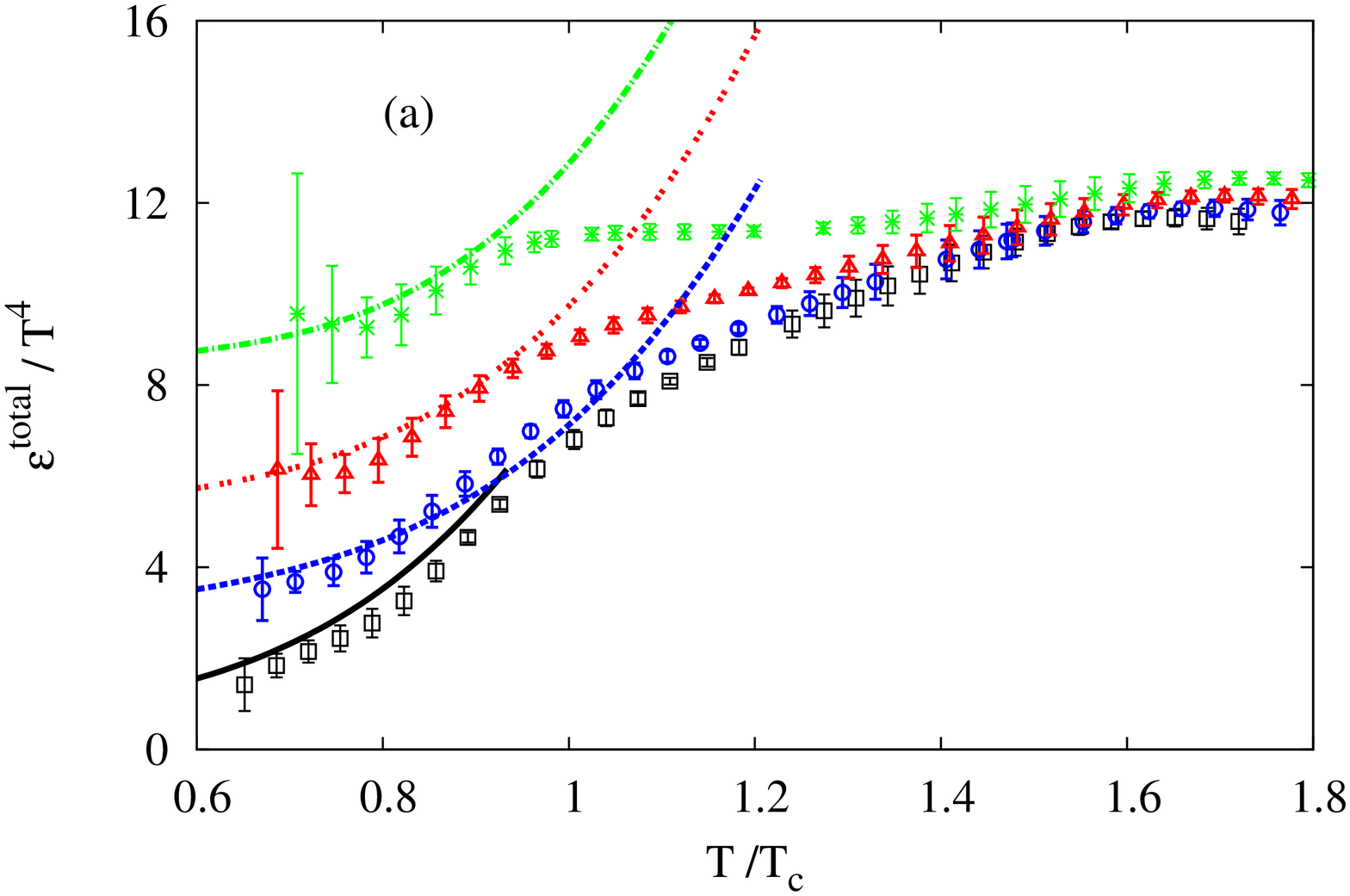}
\includegraphics[width=5.cm,angle=-90]{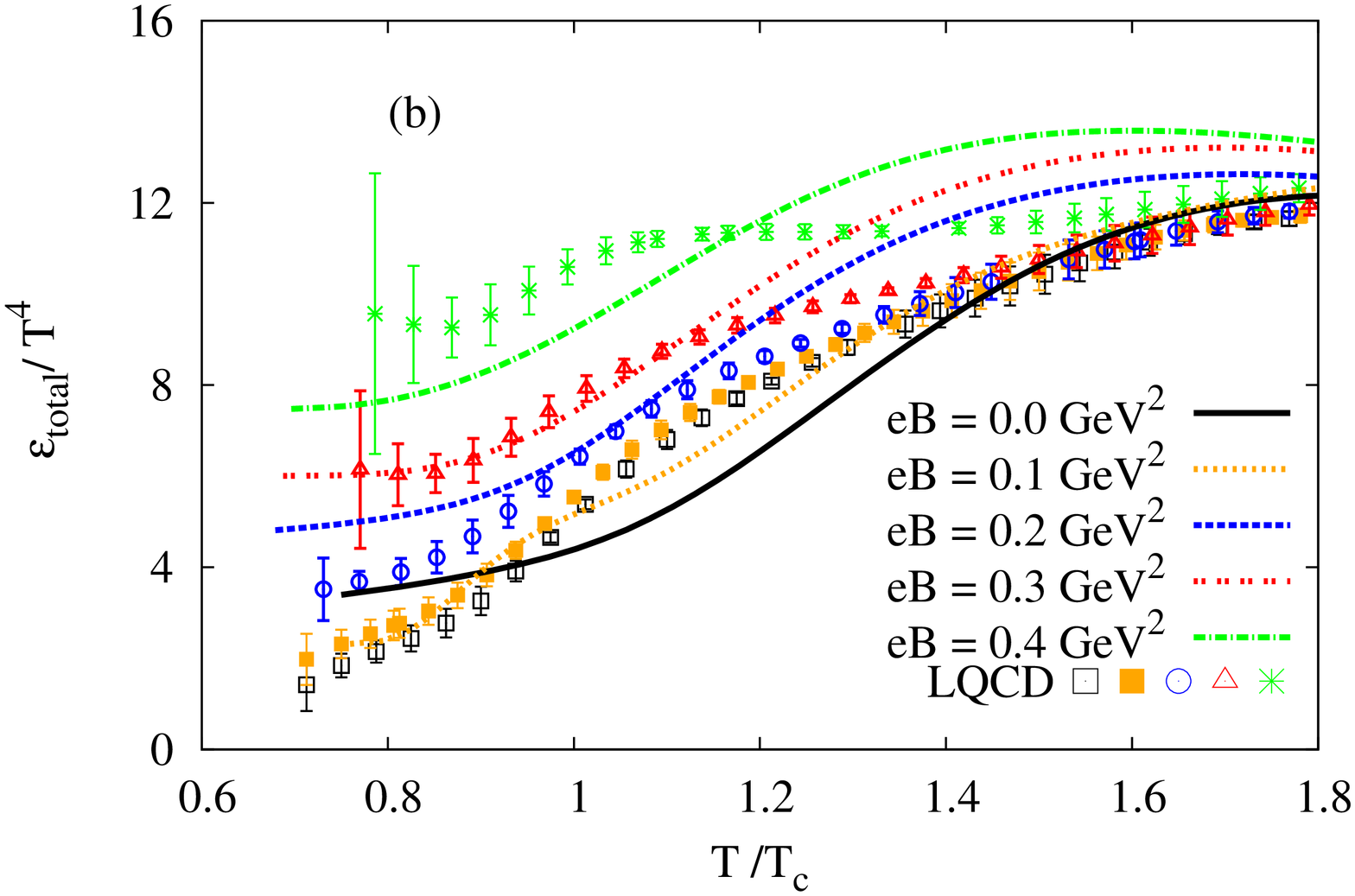} \\
\includegraphics[width=5.cm,angle=-90]{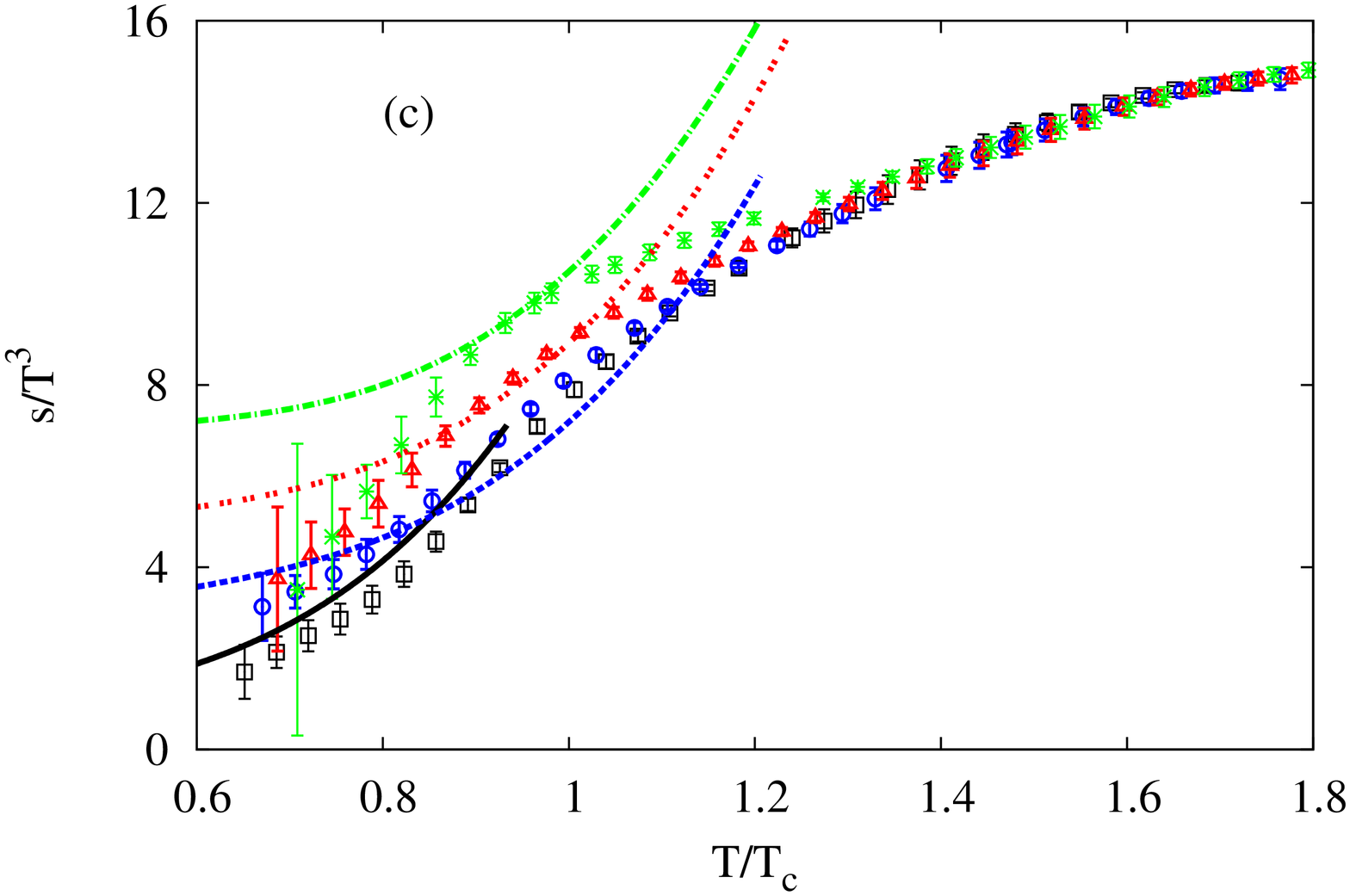}
\includegraphics[width=5.cm,angle=-90]{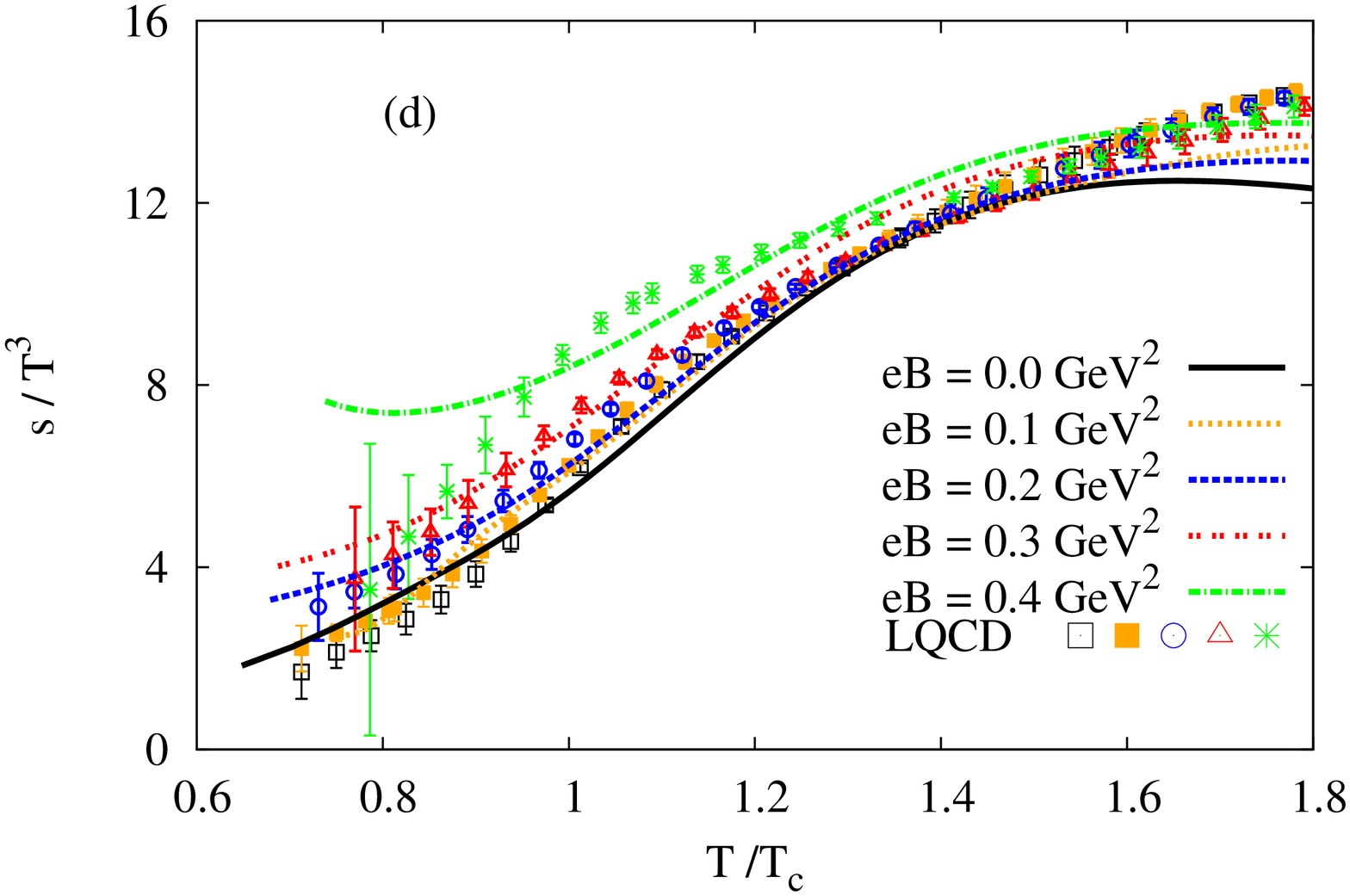}
\caption{The same as in Fig. \ref{fig:pressure} but for normalized total energy density, Eq. (\ref{eq:totalE}),  in top panels and normalized entropy density in bottom ones. }
\label{fig:entropy}
}
\end{figure}

Sign of ${\cal M}$ defines the magnetic property of the system of interest. Paramgnetic property refers to attraction of QCD matter in external magnetic field. Diamagnetic QCD matter is slightly repelled by the magnetic field and does not retain its magnetic properties when the field is removed. Positive magnetization refers to para-, while negative $\mathcal{M}$ to diamagnetism. In Fig. \ref{fig:Magnetization}, the dependence of ${\cal M}$ on temperature is depicted at various $e B$. We notice that ${\cal M}$ has very small but positive values which indicate that the QCD matter has paramagnetic property and this behavior  monotonically remains with increasing temperature. The phase transition at $T/T_c=1$ seems not affecting it. The magnetization from HRG fairly agrees with, while PLSM excellently reproduces the lattice magnetization. The slight difference in both models should contribute to further thermodynamic quantities, Fig. \ref{fig:entropy}.

In top panels of Fig. (\ref{fig:entropy}), the same as in Fig. \ref{fig:pressure} but for total energy density,
\bea
\epsilon^{total} &=& \epsilon + e\, B\cdot {\cal M}. \label{eq:totalE}
\eea
where $\epsilon=-\beta \partial p/\partial \beta$ with $\beta=1/T$, is the thermodynamic energy density. It is obvious that the agreement between HRG (a) and the lattice calculations is excellent, especially below $T_c$. Also, we observe that $T_c$ declines with increasing $e B$, i.e. inverse magnetic catalysis in the hadronic matter while $\epsilon^{total}/T^4$ remarkably increases with $e B$. In other words, we observe that increasing $e B$ leads to decreasing $T_c$. The values of $T_c$ can be deduced from the graphs. The top right-hand panel (b) shows the same but for calculations from PLSM. Here is fair agreement is extended to a wider range of temperatures covering hadronic and partonic phases, simultaneously.

The bottom panels are the same as in Fig. \ref{fig:pressure} but for entropy density $s=(\epsilon^{total}+p_z)/T$. In HRG, the increase in $s/T^3$ with increasing $e B$ is larger than in lattice calculations. The best agreement is found near $T_c$. In bottom right-hand panel (d), entropy from PLSM is confronted to the lattice simulations. There is a nice agreement in both hadronic and partonic phases.

\section{Conclusions} \label{Sec:con}

The introducing of finite magnetic field is accompanied with two types of modifications; first the phase space should be changed due to inverse magnetic catalysis and second the dispersion relation as well as the distribution function should be modified. In mean field approximation, both Polyakov linear-sigma model (PLSM) and hadron resonance gas (HRG) model are considered to study QCD thermodynamics in vanishing and finite magnetic field. The results from both models are confronted to recent lattice QCD calculations. In doing this, the temperature dependence of the magnetization should be determined. It seems that both models are fairly suited in describing the degrees of freedom in the hadronic phase, in which only PLSM contianes the partonic degrees of freedom.

In presence of finite magnetic field, both HRG and PLSM are individually compared with recent LQCD calculations at $eB=0.0-0.4~$GeV$^2$.  PLSM shows an excellent agreement with the lattice QCD results at vanishing and non-vanishing magnetic field. The construction of the PLSM plays an important role. 
The inclusion of the magnetic field can be partly modelled by changing both the dispersion relations Eqs. (\ref{disp1}) and (\ref{eqLL}) for HRG and PLSM approaches, respectively. In this doing, the momentum phase-space should be reduced as given in Eqs. (\ref{PartiHRGeB}) and (\ref{cond1}) and also scaled to the quark and hadron electric charges and the magnetic filed. This latter process is known as {\it dimension reduction} or {\it magnetic catalysis effect}. 
In inserting the magnetic field to both lattice and PLSM, the procedure implemented in is almost the same. The agreement between HRG and lattice is fairly good, especially below the critical temperature. The modified dispersion relation follows Lorentz invariance principle, where an extra term (due to magnetic energy) is added to the total energy. Accordingly, the entropy should be modified.  The sign of the magnetization depends on the magnetic property of the QCD matter. From both models and lattice calculations, we conclude that the QCD matter is likely paramagnetic. The temperature dependence is monotonic. The magnetization from HRG fairly agrees with, while PLSM excellently reproduces the lattice magnetization. The phase transition at critical temperature seems not affecting the paramagnetic property.

Furthermore, we conclude that raising the magnetic field strength increases the thermodynamic quantities, especially in the hadronic phase. At high temperatures (partonic phase), the thermodynamics is apparently limited to the Stefan-Boltzmann limits. The latter are likely slightly affected by the magnetic field strength. At high temperatures, PLSM overestimated lattice magnetization and energy density. This might be explained from the limitation of PLSM at such high temperatures. The latter can be fine-tuned while estimating chiral condensates, $\sigma_l$ and $\sigma_s$ and deconfinement order parameters, $\phi$ and $\phi^*$, respectively. The possible thermal-modifications of various thermodynamic quantities calculated from both HRG and PLSM approaches lead to the conclusion that increasing the magnetic field strength reduces the critical temperature, i.e. inverse magnetic catalysis.

\end{document}